\documentclass[aps,prl,twocolumn,showpacs,preprintnumbers,amsmath,amssymb]{revtex4-1}
\usepackage{graphicx}
\usepackage{color}
\usepackage[normalem]{ulem}  

\renewcommand\sout{\bgroup \color{red}\ULdepth=-.5ex \ULset}
\newcommand\soutb{\bgroup \color{blue} \ULdepth=-.5ex \ULset}

\usepackage{graphicx}
\usepackage{dcolumn}
\usepackage{bm}

\begin{document}

\preprint{}

\title{Sketch of the resolution of the axial U(1) problem without chiral anomaly}

\author{Nodoka~Yamanaka$^{1}$}
\affiliation{$^1$Nishina Center for Accelerator-Based Science, RIKEN, Wako 351-0198, Japan}
\email{nodoka.yamanaka@riken.jp}

\date{\today}

\begin{abstract}
We propose a mechanism which explains the masses of $\eta$ and $\eta'$ mesons without invoking the explicit violation of $U(1)_A$ symmetry by the chiral anomaly.
It is shown that the U(1) problem, the problem for which the prediction of $\eta$ and $\eta'$ masses in the simple chiral perturbation theory largely deviates from the experimental values, is actually resolved by considering the first order contribution of the disconnected meson correlator with respect to the quark mass.
The bound of Weinberg $m_\eta^2 \le 3 m_\pi^2$ is fulfilled by considering the negative squared mass of $\eta$ or $\eta'$ which is just the saddle point of the QCD effective potential, and 20\% level agreements with experimental data are obtained by just fitting one low energy constant.
We provide the leading chiral Lagrangian due to the disconnected contribution in 3-flavor QCD, and also discuss the 2- and 4-flavor cases as well as the consistency of our mechanism with the chiral restoration at high temperature found in lattice calculations.
\end{abstract}


\maketitle

It is believed that the chiral symmetry of quarks is dynamically broken in quantum chromodynamics (QCD), and that massless Nambu-Goldstone (NG) bosons appear in the spectrum \cite{Nambu:1961tp,Nambu:1961fr,Goldstone:1961eq,Goldstone:1962es}.
In reality, the quarks are massive due to explicit violation at the level of the Lagrangian so that NG particle masses are not exactly zero, but the small ``current quark masses'' nevertheless allow us to control the masses and interactions of the NG bosons in the linear approximation, which we nowadays call chiral perturbation theory \cite{Gasser:1983yg,Gasser:1984gg,Scherer:2002tk,Meissner:2024ona}.

If we consider 3-flavor QCD, one would naively expect the appearance of nine NG bosons.
However, the ninth one, $\eta'$ with its mass $m_{\eta'}= 960$ MeV, was found to be heavier than the simple prediction from chiral perturbation theory.
Moreover, this approach also predicts the existence of a light meson, $\eta$, with a mass very close to that of the pion, $m_\pi = 135$ MeV.
Even by radically changing the decay constants, the mass of $\eta$ is bounded by an inequality \cite{Weinberg:1975ui,Crewther:1978kq,Christos:1984tu}
\begin{equation}
m_\eta^2 \le 3 m_\pi^2, 
\label{eq:U(1)problem}
\end{equation}
while the observed value is $m_\eta = 550$ MeV, which seems to oppose to the simple chiral perturbation.
This is the well-known ``U(1) problem'' of Weinberg.

It is currently widely admitted that the axial U(1) symmetry [$U(1)_A$] is not conserved, as can be seen in the chiral Ward-Takahashi identity \cite{Bardeen:1969md}
\begin{equation}
\sum_{q}^{N_f}
\Bigl[
\partial^\mu (\bar q \gamma_\mu \gamma_5 q )
+2m_q
\bar q i\gamma_5 q
\Bigr]
=
-
\frac{N_f\alpha_s}{8\pi}
F_{\mu \nu, a}\tilde F^{\mu \nu}_a
,
\label{eq:chiralWTI}
\end{equation}
where the quarks $q$ are summed over $N_f$ active flavors.
The right-hand side contains the chiral anomaly $F\tilde F \equiv\frac{\alpha_s}{8\pi} F_{\mu \nu, a}\tilde F^{\mu \nu}_a$ which does not cancel even in the chiral limit \cite{Adler:1969gk,Bell:1969ts,Fujikawa:1979ay,Fujikawa:1980eg}.
A resolution to the U(1) problem was proposed by 't Hooft who claimed that the topological instanton \cite{Belavin:1975fg,Callan:1977gz,Crewther:1977ce,Schafer:1996wv}, a nonperturbative realization of the above chiral anomaly, lifts the mass of the singlet NG boson channel \cite{tHooft:1976rip,tHooft:1976snw,tHooft:1986ooh}.
It was argued that the instantons generate the $U(1)_A$ breaking 6-quark Kobayashi-Maskawa-'t Hooft (KMT) interaction \cite{Kobayashi:1970ji,Kobayashi:1971qz,Hatsuda:1994pi,Koll:2000ke}
\begin{equation}
{\cal L}_{\rm KMT}
\propto
\bar u_R u_L \bar d_R d_L \bar s_R s_L + {\rm h.c.}
.
\label{eq:KMT}
\end{equation}
As another important contribution which may shift the singlet NG boson mass, Witten and Veneziano suggested the topological susceptibility \cite{Witten:1979vv,Veneziano:1979ec,Veneziano:1980xs,Shore:2007yn,Alkofer:2008et}
\begin{eqnarray}
U(k)
&=&
\frac{2 i N_f}{f_\pi^2}
\int d^4 x\, e^{ik\cdot x}
\langle 0|
F\tilde F (x)
F\tilde F (0)
|0\rangle
,
\label{eq:topological_susceptibility}
\end{eqnarray}
which resulted in a nontrivial dependence of $\eta'$ mass on the number of color $N_c$.
Other interesting resolutions based on the gauge variance of the flavor-singlet current were also proposed \cite{Kogut:1973ab,Kugo:1978nc}.
The masses of $\eta$ and $\eta'$ were also calculated in lattice QCD, and recent results are showing good agreement with experimental data \cite{Kuramashi:1994aj,Christ:2010dd,Fukaya:2015ara,Bali:2021qem,CSSMQCDSFUKQCD:2021rvs,Verplanke:2024msi}, but it is still difficult to determine the responsible mechanism.
Until today, many subjects of QCD and hadron physics were investigated assuming the chiral anomaly, such as the hadron spectroscopy \cite{Schafer:1996wv,Oka:1989ud,Oka:2000wj,Gan:2020aco}, QCD at finite temperature and density \cite{Nagahiro:2006dr,Hatsuda:2006ps,Abuki:2010jq,Fukushima:2010bq,Jido:2011pq,Baym:2017whm,Bass:2018xmz,Suenaga:2022uqn}, and the chiral magnetic effect in accelerator experiments \cite{Kharzeev:2007jp,Fukushima:2008xe,Kharzeev:2015znc,STAR:2021pwb}.

Under the assumption of the anomalous $U(1)_A$ breaking, the resolution of the U(1) problem leads to a large CP violation due to the topological instanton effect which is in conflict with experiments, nowadays called the strong CP problem \cite{Shifman:1979if}.
However, there is recently a surge of trend that this problem is absent in QCD \cite{Ai:2020ptm,Nakamura:2021meh,Ai:2024vfa,Schierholz:2024var}, so that the U(1) problem is also getting more unstable.
In particular, the present Author recently showed that the topological charge is not observable, which immediately leads to the absence of the KMT interaction (\ref{eq:KMT}) generated by the chiral Dirac zero-modes as well as the topological susceptibility (\ref{eq:topological_susceptibility}) in the zero momentum limit $k\to 0$, and eventually to the irrelevance of the anomalous violation of global $U(1)_A$ \cite{Yamanaka:2022vdt,Yamanaka:2022bfj}.
What is then the mechanism resolving the U(1) problem?
If there is no explicit $U(1)_A$ violation by the chiral anomaly, the only source of explicit chiral symmetry breaking is then the current quark mass.
In this work, we elaborate such a mechanism and show its consistency with selected, but the most important related facts including experiments.

\begin{figure}[htb]
\includegraphics[width=6cm]{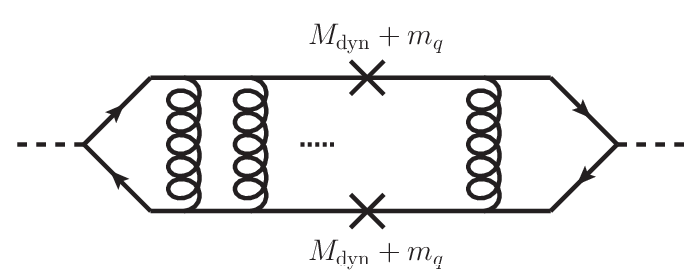}
\caption{\label{fig:meson_connected}
Connected diagram contribution to the NG boson propagator.
The solid and dashed lines denote the quark and the interpolating field of the NG boson, respectively.
The wiggle lines with the dots mean that we took the resummed diagrams with all possible gluon exchanges.
The crosses indicate the mass insertions.
}
\end{figure}

The conventional understanding of the structure of the NG bosons is as follows.
The quarks get a large mass of the order of $\Lambda_{\rm QCD} \sim 200$ MeV (the so-called ``constituent quark mass'') through the dynamics of QCD \cite{Bowman:2005vx}.
It is believed that the strong attraction due to the gluon exchange destabilizes the vacuum and forms the vacuum condensate $\langle 0 | \bar q q | 0 \rangle$ \cite{Nambu:1961tp,Nambu:1961fr,Miransky:1983vj,Higashijima:1983gx,Higashijima:1991de,Vogl:1991qt,Klevansky:1992qe,Roberts:1994dr,Hatsuda:1994pi,Eichmann:2008ae,Gubler:2018ctz}.
The zero mass of the NG bosons within massless QCD is interpreted as the exact cancellation between the dynamically generated quark mass $M_{\rm dyn}$ and the binding energy of the quark-antiquark system.
The masses of pions and kaons are due to the small current quark mass terms which disrupt the above exact cancellation.
The meson squared mass receives the leading contribution from the mass insertions $M_{\rm dyn}+m_q$ to the quark propagators of the connected meson correlator depicted in Fig. \ref{fig:meson_connected}.
To respect the chiral symmetry, we have to take an even number of mass insertions, so the leading contribution starts from $(M_{\rm dyn} + m_q)^2$, but we also have to subtract the $M_{\rm dyn}^2$ term due to the NG boson binding.
This then yields a formula where the NG boson mass squared is linear in the current quark mass, as expected.

We now formulate more quantitatively the squared NG boson mass by respecting the $U(3)_L \times U(3)_R$ symmetry.
In the $SU(3)$ nonet subbasis $[\eta_3 (\equiv \pi_0) , \eta_8 , \eta_0]$, we have \cite{Weinberg:1975ui}
\begin{eqnarray}
&&
\hspace{-2em}
M_\phi^2
=
\nonumber\\
&&
\hspace{-2em}
\left(
\begin{array}{ccc}
2 m_l B_{33} &\frac{1}{\sqrt{3}} m_- B_{38} & \sqrt{\frac{2}{3}} m_- B_{30} \cr
\frac{1}{\sqrt{3}} m_- B_{38} & \frac{2}{3} (m_l + 2 m_s) B_{88} & \frac{2\sqrt{2}}{3} (m_l - m_s) B_{80} \cr
\sqrt{\frac{2}{3}} m_- B_{30} & \frac{2\sqrt{2}}{3} (m_l - m_s) B_{80} & \frac{2}{3} (2m_l + m_s) B_{00}  \cr
\end{array}
\right)
,
\nonumber\\
\label{eq:NGmassmatrix1}
\end{eqnarray}
where the quark masses are given by $m_l \equiv \frac{1}{2} (m_u + m_d)= 3.43$ MeV, $m_- \equiv (m_u - m_d)= -2.54$ MeV, and $m_s = 93.5$ MeV (at the renormalization scale $\mu = 2$ GeV) \cite{ParticleDataGroup:2024cfk}, respectively.
The linear coefficients $B$ are defined by $B_{ij} =- \frac{\langle 0 | \bar q q | 0 \rangle}{f_i f_j}$ $(i,j=0,3,8)$, where $f_3 \equiv f_\pi = 92.2$ MeV, $f_8 = 115$ MeV, and $f_0 = 100$ MeV \cite{Bali:2021qem}.
Fitting the squared pion mass $m_\pi^2 =2 m_l B=(135\, {\rm MeV})^2$, we obtain $B_{33} \approx 2.6$ GeV and $\langle 0 | \bar q q | 0 \rangle \approx -(280 \, {\rm MeV} )^3$.
Here we assumed that $\langle 0 | \bar q q | 0 \rangle = \langle 0 | \bar uu | 0 \rangle = \langle 0 | \bar dd | 0 \rangle = \langle 0 | \bar ss | 0 \rangle$ \cite{Gubler:2018ctz,McNeile:2012xh}.
The diagonalization of the matrix (\ref{eq:NGmassmatrix1}) yields $m_\pi \approx 148$ MeV, $m_\eta \approx 98$ MeV, and $m_{\eta'} \approx 589$ MeV, with the $\eta - \eta'$ mixing angle $\theta \approx -51^{\circ}$.
We obtained a very light $\eta$ meson having a comparable mass to the pion which is in line with Weinberg's U(1) problem \cite{Weinberg:1975ui}.
We emphasize that tuning the decay constants $f_0 , f_8$ so as to fit $m_{\eta'}$ cannot resolve the issue of small $m_\eta$.

Here we remind the readers of the derivation of the U(1) problem (\ref{eq:U(1)problem}).
Weinberg considered the isoscalar meson with the state $\phi = (0 , f_8  , \sqrt{2} f_0) /\sqrt{3}$ (again in the nonet subbasis).
He remarked that any expectation value of a positive mass matrix is less than its minimal eigenvalue, which lead to \cite{Weinberg:1975ui}
\begin{equation}
m_L^2 \phi^\dagger \phi = \frac{1}{3} m_L^2 (f_8^2+ 2f_0^2)
\le
\phi^\dagger M_\phi^2 \phi
= f_\pi^2 m_\pi^2 
,
\end{equation}
where $m_L^2$ is the lowest eigenvalue of the matrix (\ref{eq:NGmassmatrix1}), corresponding to the squared mass of $\eta$.
By taking $f_8=f_\pi$ and $f_0 =0$, we find the most extreme inequality (\ref{eq:U(1)problem}).

We actually have an alternative understanding of the U(1) problem.
If we take the approximation of $m_s \gg m_u , m_d$ and $B_{00}\approx B_{80}\approx B_{88}$, the mass matrix (\ref{eq:NGmassmatrix1}) becomes
\begin{eqnarray}
M_\phi^2
&\propto &
\left(
\begin{array}{ccc}
0 & 0& 0 \cr
0 & 2 & -\sqrt{2} \cr
0 & -\sqrt{2} & 1 \cr
\end{array}
\right)
.
\label{eq:NGmassmatrix1.5}
\end{eqnarray}
We then see that the second and third line vectors are collinear, so this matrix is just rank-1 and there is only one nonzero eigenvalue.
This means that the mass matrix (\ref{eq:NGmassmatrix1}) can only yield one heavy state, and that the other two are as light as the light pion generated by $m_u$ and $m_d$, neglected in Eq. (\ref{eq:NGmassmatrix1.5}).
This is actually another expression of Weinberg's U(1) problem.
To obtain a heavy $\eta$, it is then mandatory to destroy the collinearity between the second and third lines of the matrix (\ref{eq:NGmassmatrix1.5}).

\begin{figure}[htb]
\includegraphics[width=6cm]{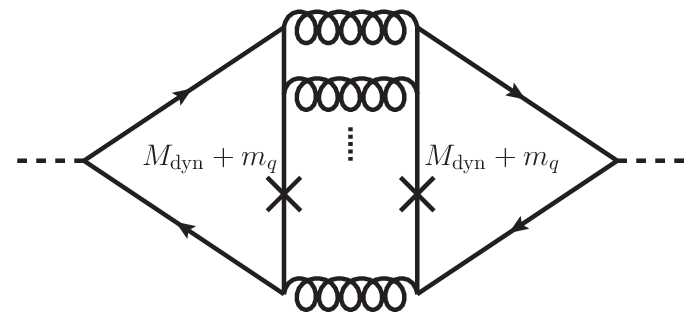}
\caption{\label{fig:meson_disconnected}
Disconnected diagram contribution to the propagator of the flavor singlet NG boson.
We use the same definitions as Fig. \ref{fig:meson_connected}.
This can effectively be viewed as an off-shell t-channel propagation of an NG boson between the quark and the antiquark.
}
\end{figure}

The conventional resolution of the U(1) problem was so far to add a flavor singlet term as
\begin{eqnarray}
&&
M_\phi^2
+
\left(
\begin{array}{ccc}
0 & 0& 0 \cr
0 & 0& 0 \cr
0 & 0& U \cr
\end{array}
\right)
,
\label{eq:NGmassmatrix2}
\end{eqnarray}
where $U \approx (860\, {\rm MeV})^2$.
This trivially breaks the collinearity between the second and third lines of the matrix (\ref{eq:NGmassmatrix1.5}).
After diagonalization of Eq. (\ref{eq:NGmassmatrix2}), we obtain $m_\pi \approx 133$ MeV, $m_\eta \approx 418$ MeV, and $m_{\eta'} \approx 959$ MeV, which are in good agreement with known data, with the $\eta - \eta'$ mixing angle $\theta \approx -13^{\circ}$.
However, we saw that there are no candidates of $U$ if we give up with the fact that the anomalous violation of $U(1)_A$ is physical \cite{Yamanaka:2022vdt,Yamanaka:2022bfj}.
Here we propose that an additional contribution due to the disconnected diagram \cite{DeRujula:1975qlm} can resolve the problem.
The leading disconnected contribution to the meson correlator is shown in Fig. \ref{fig:meson_disconnected}.
To not cancel the trace of Dirac matrices, we need at least one mass insertion for each quark loops, yielding a factor of $(M_{\rm dyn} + m_q)^2$.
Here we see that the gluon exchange between quark loops can take the configuration of an NG boson propagating between the quarks in the t-channel, just like the meson exchange in the quark-meson model.
We therefore again have the NG boson binding which removes the constituent quark contribution $M_{\rm dyn}^2$, and the leading order contribution to the disconnected diagram again becomes linear in $m_u , m_d$, and $m_s$.
The mass matrix should have a form like
\begin{eqnarray}
&&
M_{\rm disc}^2
=
\nonumber\\
&&
\left(
\begin{array}{ccc}
0 & 0& \sqrt{\frac{3}{2}} m_- B'_{30} \cr
0 & 0& \sqrt{2} (m_l -m_s) B'_{80} \cr
\sqrt{\frac{3}{2}} m_- B'_{30} & \sqrt{2}(m_l -m_s) B'_{80} & 2 (2m_l+m_s) B'_{00} \cr
\end{array}
\right)
.
\nonumber\\
\label{eq:NGmassmatrix3}
\end{eqnarray}
where the low energy constant is defined as $B'_{ij} \equiv B_{\rm disc} \frac{f_\pi^2}{f_i f_j }$ ($i,j= 3,8,0$).
This contribution was already noticed as a higher order contribution in the $1/N_c$ expansion \cite{Christos:1982rp,Christos:1984tu}, but it was never analyzed under the absence of the chiral anomaly effect.
The second and the third lines of the above disconnected contribution are clearly linearly independent, so it is possible to avoid the U(1) problem.

\begin{figure}[htb]
\includegraphics[width=4cm]{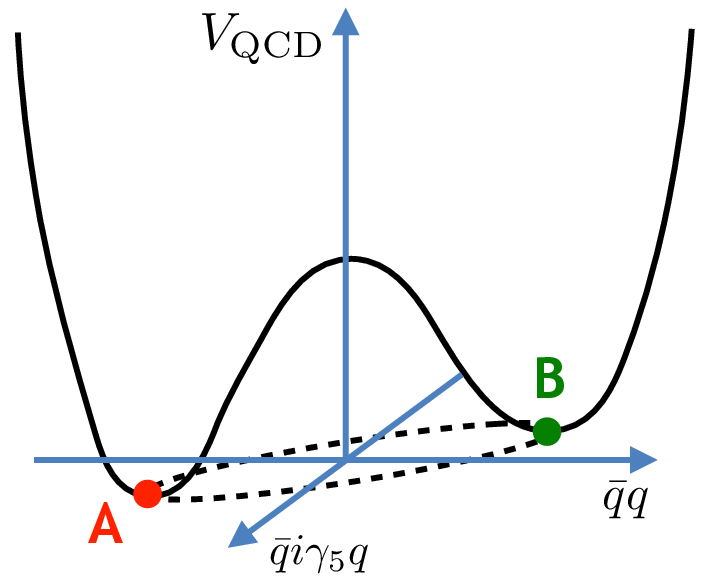}
\caption{\label{fig:QCD_potential}
The schematic picture of the effective potential of QCD in the function of quark scalar and pseudoscalar bilinears.
The dashed line denotes the tilted chiral circle due to the current quark mass.
The red point A represents the absolute minimum with a positive squared mass, while the green point B is the unstable saddle point with a negative curvature in the direction of the chiral circle.
The diagonalization of the meson squared mass matrix $M_\phi^2 + M_{\rm disc}^2$ yields the latter for either $\eta$ or $\eta'$.
}
\end{figure}

We now fit the low energy constant $B_{\rm disc}$ of Eq. (\ref{eq:NGmassmatrix3}) so as to reproduce either $m_\eta$ or $m_{\eta'}$.
The resulting possibilities obtained by diagonalizing $M_\phi^2 +M_{\rm disc}^2$ are displayed in Table \ref{table:Bdiscfit}.
We remark that one of the two mass eigenvalues is negative.
One might think that these negative squared masses are unphysical, but this is actually not the case.
Let us remind that the squared NG boson mass was linear in the current quark masses $m_u , m_d$, and $m_s$.
If we draw the effective potential of QCD in terms of the quark scalar and pseudoscalar bilinears, the current quark mass just gives the tilt to the wine bottle potential, as depicted in Fig. \ref{fig:QCD_potential}.
It is a well-known fact that the NG bosons are the excitation of the quark field bilinear in the direction of the chiral phase (also called ``chiral circle''), and in massless QCD, this angular direction is totally flat so that the excitation has no cost, which explains the zero mass of the NG boson.
As mentioned above, the potential is tilted when quarks have masses in the QCD Lagrangian, and the angular direction also obtains a minimum, where the curvature is positive (point A of Fig. \ref{fig:QCD_potential}).
In the opposite side of the chiral circle we also have a saddle point, where the curvature of the angular direction is negative (point B of Fig. \ref{fig:QCD_potential}).
This is exactly the negative squared mass that was needed to explain the results of Table \ref{table:Bdiscfit}.
The negative squared NG boson mass just means that there is another true minimum in the opposite side of the saddle point.
We do not know the precise form of the QCD potential, but by assuming that the tilt is not important compared to the curvature in the radial direction, the negative and positive masses should approximately have the same magnitude.
We can then just revert the signs of the negative squared masses of Table \ref{table:Bdiscfit}.
We see that Fits III and IV reproduce well the observed experimental values of $m_{\eta}$ and $m_{\eta'}$, and the deviations are in both cases less than 10\%.
For them, the low energy constant $B_{\rm disc}$ is negative, which is consistent with the fact that the disconnected diagram is having an extra quark loop bringing a factor of $-1$ relative to the connected one (Fig. \ref{fig:meson_disconnected}).
The mixing angle $\theta \sim -18^\circ$ obtained from Fits III and IV also agrees well with the phenomenology \cite{Gilman:1987ax,Chao:1989yp,Moussallam:1994xp,Ball:1995zv,Herrera-Siklody:1997pgy,Bass:2018xmz,Gan:2020aco} and also with lattice results \cite{Christ:2010dd}.
To see the success of the fits, we can for instance calculate the 2-photon decay rates of $\eta$ and $\eta'$ with $\theta$ as input.
In Table \ref{table:decay}, we show the values obtained from our results in the leading order analysis \cite{Rosner:1982ey}.
We see that the mixing angle of Fit IV reproduces well the experimental data, within 20\%.

\begin{table}
\caption{
Different possible fits of $B_{\rm disc}$.
The $\eta - \eta'$ mixing angle is denoted by $\theta$.
The pion mass is in any cases $m_\pi \approx 135$ MeV.
}
\begin{ruledtabular}
\begin{tabular}{c|cccc}
Fit no. & $B_{\rm disc}$ (GeV) & $m_\eta^2$ (MeV$^2$) & $m_{\eta'}^2$ (MeV$^2$) & $\theta$ ($^\circ$) \\
\hline
I & 9.00 & -(550)$^2$ & (1483)$^2$ & -27 \\
II & 2.92 & -(246)$^2$ & (958)$^2$ & -32 \\
\hline
III & -4.91 & (550)$^2$ & -(872)$^2$ & -17 \\
IV & -5.58 & (573)$^2$ & -(959)$^2$ &  -18 \\
\end{tabular}
\end{ruledtabular}
\label{table:Bdiscfit}
\end{table}

\begin{table}
\caption{
Comparison of predicted 2-photon decay rates and their experimental values.
All values are in unit of eV.
}
\begin{ruledtabular}
\begin{tabular}{l|cccc}
$\Gamma$ & $\theta = -13^\circ$ & $\theta = -32^\circ$ & $\theta = -18^\circ$ & Exp. \cite{ParticleDataGroup:2024cfk}  \\
\hline
$\eta \to \gamma \gamma $ & 320 & 729 & 418 & 516(20) \\
$\eta' \to \gamma \gamma $ & 5170 & 2970 & 4640 & 4340(140) \\
\end{tabular}
\end{ruledtabular}
\label{table:decay}
\end{table}

The chiral effective interaction that can be matched with the disconnected contribution (\ref{eq:NGmassmatrix3}) is for instance \cite{DiVecchia:1980vpx,Herrera-Siklody:1996tqr,Herrera-Siklody:1997pgy,Kaiser:2000gs}
\begin{equation}
{\cal L}_{\rm disc}
=
\frac{i\sqrt{3}}{2\sqrt{2}}
B_{\rm disc} f_\pi
\eta_0 
{\rm Tr} 
\Biggl[ 
M_q 
\exp \Biggl(
\sum_{a=0}^{8}
\frac{i\lambda_a \eta_a}{f_\pi} 
\Biggr)
+{\rm h.c.}
\Biggr]
,
\label{eq:chiral_lagrangian}
\end{equation}
where $M_q \equiv {\rm diag}[m_u, m_d, m_s]$, $\lambda_i$ ($i=1,8$) are the Gell-Mann matrices, and $\lambda_0 \equiv {\rm diag}[1,1,1] / \sqrt{6}$.
The decay constant is $f_\pi$ in the leading order, and $\eta_a$'s are the NG bosons defined in the nonet basis.
Since the chiral anomaly plays no more central role in the chiral perturbation, we can inspect processes involving $\eta$ and $\eta'$ in the ordinary chiral perturbation without caring the $1/N_c$ expansion, such as $\eta \to 3 \pi$ which has been extensively studied in the past \cite{Osborn:1970nn,Kogut:1974kt,Kawarabayashi:1980dp,Abdel-Rehim:2002fen,Bijnens:2002qy,Bijnens:2007pr,Guo:2015zqa,Albaladejo:2017hhj,Bass:2018xmz,Gan:2020aco,BESIII:2023edk}.
Detailed analyses are left as future works.

Let us try to understand how the U(1) problem is critical in QCD from the point of view of $N_f$ and the quark masses.
In the $N_f = 2$ case, the connected and disconnected contributions would look like
\begin{eqnarray}
M_\phi^2
&=&
\left(
\begin{array}{cc}
2m_l & m_- \cr
m_- & 2m_l \cr
\end{array}
\right)
B
,
\\
M_{\rm disc}^2
&=&
\left(
\begin{array}{cc}
0 & m_- \cr
m_- & 4 m_l \cr
\end{array}
\right)
B_{\rm disc}
.
\label{eq:Nf=2}
\end{eqnarray}
The diagonalization of $M_\phi^2$ (connected) just yields the $\bar{u}u$ and $\bar{d}d$ NG bosons as a result of the ideal mixing, with their masses $m_{\bar{u}u} = \sqrt{2B m_u} = 106$ MeV and $m_{\bar{d}d} = \sqrt{2B m_d} = 156$ MeV.
We see that there is no Weinberg's U(1) problem for $N_f =2$ since there is no large hierarchy between $m_u$ and $m_d$, 
Now let us add $M_{\rm disc}^2$.
By using $B_{\rm disc}=-5.58$ GeV (Fit IV of Table \ref{table:Bdiscfit}), 
we obtain $m_\pi = 134$ MeV and $m_\eta = 244$ MeV, where we have again removed the minus sign of $m_\eta^2$ appearing after the diagonalization of $M_\phi^2 +M_{\rm disc}^2$ (saddle point of Fig. \ref{fig:QCD_potential}).
The inspection of the quark mass dependence of $\pi$ and $\eta$ mesons in $N_f = 2$ QCD in lattice calculation may then validate our statement in the future.
The above results were obtained by assuming that $B$ and $B_{\rm disc}$ have no $N_f$-dependence.
For the $N_f = 3$ case studied in this work, the light quark masses have little influence on the masses of $\eta$ and $\eta'$ due to the mass hierarchy $m_s \gg m_u , m_d$.
We numerically checked that they are within a few \%.

For the 4-flavor case, we have a very large mixing between the ninth generator of SU(4) $t_{9} = {\rm diag} [1,1,1,-3]/\sqrt{6}$ (corresponding to $\eta_0$ in $N_f =3$) and the SU(4) singlet due to the heavy charm quark $m_c \approx 1.27$ GeV $\gg M_{\rm dyn} $.
The leading contribution to the meson mass starts from $O(m_c^2)$, and we have an ideal mixing just like Eq. (\ref{eq:NGmassmatrix1.5}), so that the charmed contribution completely splits from the other quarks.
The $\eta_c$ Fock state is then almost composed of charm and anticharm quarks.
From the above inspection, we arrive at the conclusion that the U(1) problem arose because of the mass hierarchy $ m_c \gg M_{\rm dyn} > m_s \gg m_u , m_d$ and the analysis can be restricted to $N_f =3$ as expected.

Let us finally comment on the finite temperature transition of QCD.
At finite temperature, the effective thermal masses of quarks and gluons adds to the QCD potential.
This will then just increase the curvature of the potential near the origin in Fig. \ref{fig:QCD_potential}, and we will end up with a potential convex downward at sufficiently high temperature.
Therefore $U(1)_A$ is, like other symmetries, restored at high temperature, as suggested by recent lattice results \cite{Cossu:2013uua,Brandt:2016daq,Aoki:2020noz}.
We emphasize that this is consistent with the irrelevance of the explicit $U(1)_A$ breaking by the chiral anomaly \cite{Yamanaka:2022vdt,Yamanaka:2022bfj}.
Since the QCD potential is tilted due to the non-vanishing quark masses, its absolute minimum will never reach the origin, so the QCD transition at finite temperature is a crossover.

In summary, we found a possible resolution of the U(1) problem, i.e. the derivation of the masses of $\eta$ and $\eta'$, without considering the explicit breaking of $U(1)_A$ due to the chiral anomaly.
This result is consistent with the earlier statement of Refs. \cite{Yamanaka:2022vdt,Yamanaka:2022bfj} that the topological charge of QCD is unphysical, and also with finite temperature lattice analyses just seen above.
We also found a nontrivial interpretation of the negative squared mass of the NG bosons, which was crucial to avoid Weinberg's U(1) problem.
Our analysis suggests that $\eta$ and $\eta'$ also obey ``normally'' the chiral perturbation theory, and that the expansion in terms of the strange quark mass is not bad.
We could also have a hint via the diagrammatic analysis that the NG bosons are just bound states of ``constituent quarks'' dressed by gluons and that their leading dynamical mass effect was removed by the binding energy due to the confining potential.
This picture fully agrees with the quark model.
To consolidate our scenarios, the dynamics of processes involving $\eta$ and $\eta'$ must also be analyzed with the fitted chiral Lagrangian (\ref{eq:chiral_lagrangian}).
This is left as future works.

\begin{acknowledgments}
The Author thanks Makoto Oka for useful discussions and comments.
This work (project) was supported by the RIKEN TRIP initiative (Nuclear transmutation), and also partially by ERATO project-JPMJER2304.
\end{acknowledgments}

\end{document}